\documentclass[aps, prl, reprint, amsmath, amssymb]{revtex4-1}
\usepackage[colorlinks,allcolors=blue]{hyperref}
\usepackage{bm}

\usepackage[english]{babel}
\usepackage{color}
\usepackage{xcolor}
\usepackage{xfrac}
\usepackage{graphicx}

\DeclareMathAlphabet{\mathsfit}{\encodingdefault}{\sfdefault}{m}{sl}
\SetMathAlphabet{\mathsfit}{bold}{\encodingdefault}{\sfdefault}{bx}{sl}

\newcommand{\gamd}{\ensuremath{{\dot{\gamma}}}}

\newcommand{\pe}{\ensuremath{\mathrm{Pe}}}

\newcommand{\pes}{\ensuremath{\mathrm{Pe}_\sigma}}

\begin{document}
\textbf{Reply to the comment from Ikeda, Berthier, and Sollich (IBS)~\cite{Ikeda_comment-yieldglass_prl2016}}

We thank IBS for their comments which question our interpretation of the universal viscosity divergence near the flow-arrest transition in constant stress and pressure rheology of hard-sphere colloidal suspensions~\cite{Wang_const-stress-pressure_prl2015}.  IBS introduced two P\'{e}clet numbers: $\pe_0 = \gamd a^2/d_0$ and $\pe=\gamd a^2/d(\phi)$, with $\gamd$ the strain rate, $a$ the particle size,  $d_0$ the isolated single-particle diffusivity and $d(\phi)$ the long-time at-rest self-diffusivity, and considered three regimes: (i) $\pe_0<\pe\ll 1$, (ii) $\pe_0 \ll 1 \ll\pe $, and (iii) $1 \ll \pe_0<\pe$.

IBS's claim that ``only $\pe$ is considered in \cite{Wang_const-stress-pressure_prl2015}'' is not true.  The stress P\'{e}clet number $\pes = \sigma a^2/(\eta_0 d_0)$, with $\sigma$ the imposed stress and $\eta_0$ the solvent viscosity, is a primitive input to our simulations.  It compares the magnitude of the imposed stress relative to the particle thermal fluctuations, and is trivially connected to $\pe_0$ through $\pes=\eta\pe_0$, with $\eta$ the dimensionless shear viscosity.

\begin{figure}[ht]
  \centering
  \includegraphics[width=3in]{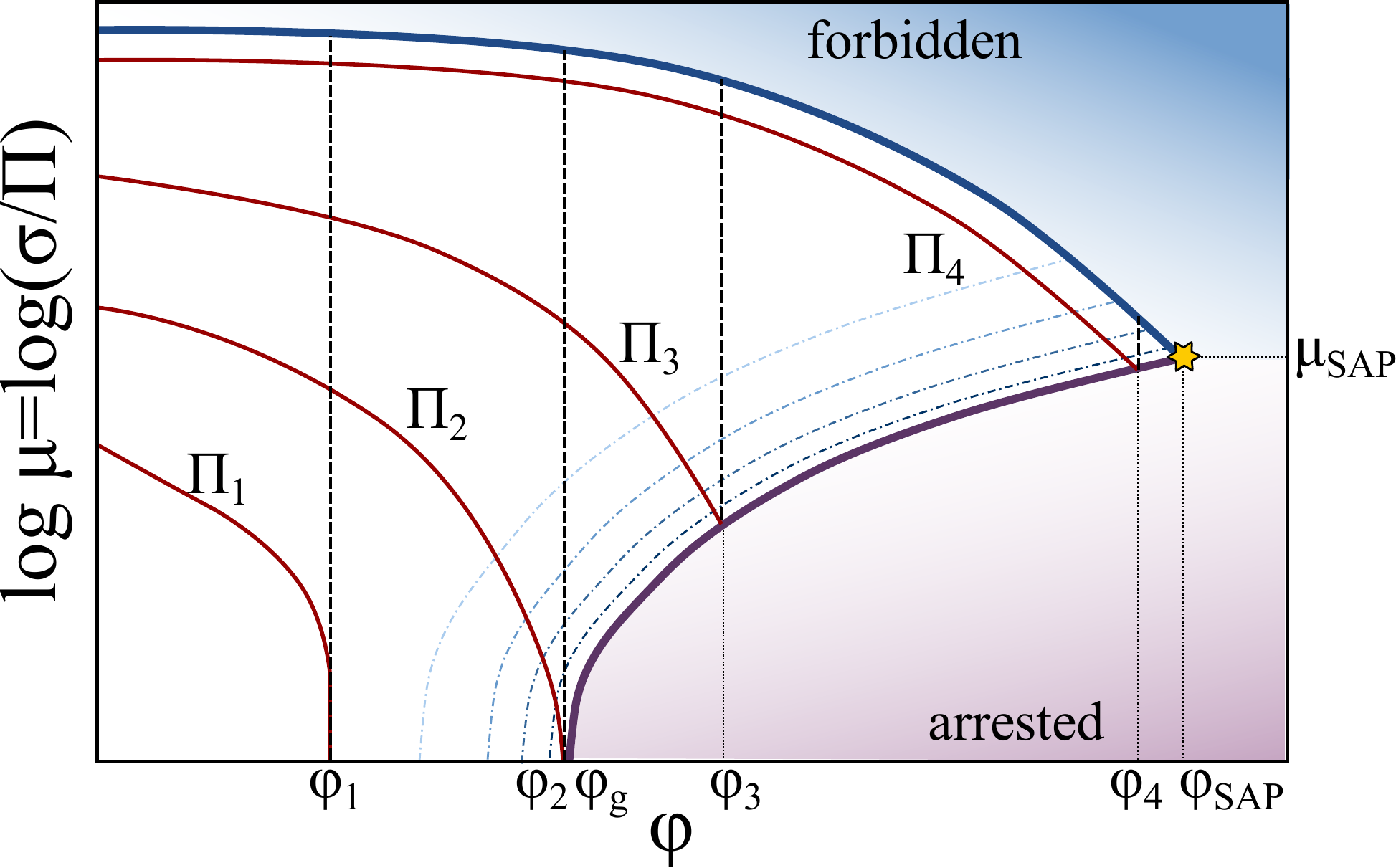}
  \caption{\label{fig:sketch}
Sketch of the $\mu$-$\phi$ flow map based on \cite{Wang_const-stress-pressure_prl2015}.  The thick curves enclose the flowing region, with the lower curve outlining the arrested region, and the upper curve outlining the non-Brownian limiting behavior.  The two curves intersect at the Shear Arrest Point $(\mu_{\mathrm{SAP}}, \phi_{\mathrm{SAP}})$.  The solid lines represent constant-$\Pi$ contours at pressures $\Pi_1<\Pi_2<\Pi_3<\Pi_4$.  The dashed lines show the constant-$\phi$ contours at the corresponding at-rest volume fraction.  The dash-dotted lines are the constant-viscosity contours. }
\end{figure}

Near the flow-arrest transition, $\pe_0$ is of little relevance to suspension dynamics.  What drives an otherwise arrested suspension to flow are  internal structural rearrangements, which are characterized by $d(\phi)$, not by the local ``in cage'' thermal fluctuations described by $d_0$.  Near athermal jamming, i.e., close to the point $(\phi_\mathrm{SAP}, \mu_\mathrm{SAP})$ in Fig.~\ref{fig:sketch}, the condition $\pe_0\gg 1$ is not satisfied.  Here, the imposed pressure $\bar{\Pi} = \pes/(6\pi \mu_{\mathrm{SAP}})$ satisfies $\bar{\Pi}\sim(\phi_{\mathrm{SAP}}-\phi)^{-\delta}$ with $\delta = 1$ near jamming~\cite{hs_metastability_crystal_torquato_prl96}.  Meanwhile, the universal viscosity divergence suggests $\pe_0\sim \pes (\phi_{\mathrm{SAP}}-\phi)^\gamma$ with $\gamma\approx2$, which leads to $\pe_0 \sim \mu_{\mathrm{SAP}} (\phi_{\mathrm{SAP}}-\phi)^{\gamma-\delta}$, independent of $\pes$ and $\bar{\Pi}$.  Thus, $\pe_0\ll 1$ for $\gamma>\delta$, which is the case for hard-sphere suspensions when $(\pes, \phi)\rightarrow(\infty, \phi_{\mathrm{SAP}})$.  IBS's distinction between regimes (ii) and (iii) is therefore unnecessary, and $\pe$ alone is sufficient.  This is also reflected in recent experiments~\cite{Poon_rheo-hs-particle_PRL2015} which show that suspensions enter the non-Brownian regime sooner, i.e., at lower $\pe_0$, with increasing $\phi$---the shear stresses where the shear thinning regime ends are the same over a wide range of $\phi$.

In regime (i), linear response theory requires $\Pi(\phi, \gamd)=\Pi^{\mathrm{eq}}(\phi) + \Delta\Pi(\phi)\gamd^2$ and $\sigma(\phi, \gamd) =\eta_T(\phi)\gamd $.  Due to the different $\gamd$ dependences, one can always evaluate $\eta_T(\phi)$ at sufficiently small $\gamd$ with $\Pi\approx\Pi^{\mathrm{eq}}(\phi)$.  In the low $\mu$ limit,  constant $\Pi$ and  constant $\phi$ results are equivalent.  This is shown in Fig.~\ref{fig:sketch}: Far from the glass transition $\phi_g$, the contour at constant $\Pi_1$ asymptotes to the contour at constant $\phi_1$ at a low but finite $\mu$.  Near $\phi_g$, the contours at $\Pi_2$ and $\phi_2$ approach each other  as $\mu\rightarrow 0$.  Therefore, by construction, our approach can probe the glass transition.  On the other hand, the viscosity divergences observed along constant-$\phi$ and constant-$\Pi$ contours may be different due to the different approaches to the arrested region~\cite{Wang_const-stress-pressure_prl2015}, as illustrated by the viscosity contours in Fig.~\ref{fig:sketch}.  Furthermore, it is still an open question whether the product $\eta_T(\phi) d(\phi)$ remains constant near $\phi_g$, and, consequently, simulations and experiments of the relaxation time~\cite{hs-glass_brambilla_prl09} cannot infer  the viscosity divergence~\cite{Ikeda_comment-yieldglass_prl2016}.

When the at-rest volume fraction is above $\phi_g$, the diffusivity $d(\phi)\rightarrow 0$ and the suspension has a yield stress.  This corresponds to IBS's regimes (ii) and (iii).  Here, the viscosity is inherently non-Newtonian regardless of $\pe_0$, and exhibits universal divergences at constant $\Pi$.  IBS's interpretation using a Herschel-Bulkey model for the pressure nicely complements our work.  Our study is for true hard spheres whose behavior can be fundamentally different from soft-particle systems, even when the stiffness of the potential is increased~\cite{jam-glass-rheo-soft-part_ikeda_prl2012} or the confining pressure is reduced~\cite{Berthier_div-visc-non-brown-susp_PRE2015} to eliminate particle overlaps.  For example, in the non-Brownian limit, the singular hard-sphere potential leads to a finite shear viscosity despite the stress's thermal origin~\cite{BRAjfm97,*BERjfm02}.  In the same limit, the viscosity from a soft potential (no matter how stiff) approaches zero.

Finally, we agree with IBS that in their regime (iii), our data are sparse since $\phi_{\mathrm{SAP}}$ can only be approached from below in our simulations.  However, as we have already pointed out, $\pe_0\gg 1$ cannot be achieved near athermal jamming, and our results agree with the viscosity divergence found in non-Brownian experiments~\cite{boyer-granular-rheology_prl2011}.\\

M. Wang${}^{1}$ and J. F. Brady${}^{1}$\\

${}^{1}$Division of Chemistry and Chemical Engineering, California Institute of Technology, Pasadena, CA 91125, USA


%

\end{document}